# A photon-level broadband dual-comb interferometer

# for turbulent open-air trace gases detection application


Wei Zhong[1], Yingyu Liu[1], Qin Yin[1], Ruocan Zhao[1,2,4], Yiwei Ding[1], Chong Wang[1,2,4], Tindi Chen[1,2,4], Xiankang Dou[4,*] & Xianghui Xue [1,2,3,4,*]

[1]CAS Key Laboratory of Geospace Environment, School of Earth and Space Sciences, University of Science and Technology of China, Hefei 230026, China.

[2]CAS Center for Excellence in Comparative Planetology, Anhui Mengcheng Geophysics National Observation and Research Station, University of Science and Technology of China, Hefei, China

[3]Hefei National Laboratory for Physical Sciences at the Microscale and School of Physical Sciences, University of Science and Technology of China, Hefei 230026, China

[4]Hefei National Laboratory, University of Science and Technology of China, Hefei 230088, China

*Corresponding authors: xuexh@ustc.edu.cn


## Priprint at arXiv. Jan 20,2024


**Abstract:** Open-path dual-comb spectroscopy (DCS) significantly enhances our understanding of regional trace gases. However, due to technical challenges, cost considerations, and eye-safety regulations, its sensing range and flexibility remain limited. The photon-counting DCS demonstrated recently heralds potential innovations over open-path DCS. Nevertheless, a major challenge in open-air applications of this approach lies in accurately extracting information from the arrival time of photons that have traversed the turbulent atmosphere. Here, we demonstrate a photon-level dual-comb interferometer for field deployment in open-air environments, uniquely designed to counteract the impact of optical path-length variations caused by atmospheric turbulence and fiber-length wandering. Under variable optical path-length conditions, 20nm broadband absorption spectrum of $H^{13}C^{14}N$ is acquired, with the power per comb line detected as low as 4 attowatt —$10^{10}$ times more sensitive than traditional DCS. Furthermore, this photon-






level DCS achieves comb-line resolution with a quantum-noise-limited signal-to-noise (SNR). This paves the way for novel open-path DCS applications, including non-cooperative target sensing and sensing over a hundred-kilometers range, all within a portable, fieldable, eye-safety and low power consumption system.

## Introduction

Trace gases contributing to environmental pollution and global warming, are increasingly drawing concern, practically in relation to their mole fractions distributions and fluxes variations at a city scale. An ideal open-path broadband spectroscopy sensor can identify the regional sinks, sources and transportations of multiple trace gases[1]. It is a good complementary to in-situ sensors and satellite sensing by covering localized areas spanning several meters and vertical columns extending to hundreds of kilometers, respectively. Among several open-path technologies available such as open-path FTR [1,3], LP-DOAS [4,5], and open-path TDLAS [6], there is a pressing need to address issues of precision, reproducibility and limited path length. To this end, open-path DCS has emerged over the last decade as a powerful tool to overcome these challenges and study multiplies trac e gases [7-18]. A proof of principle study in 2014 [7] has demonstrated the ability to retrieve dry concentrations of carbon dioxide and methane over 1km path, achieving a precision of less than 1ppm for $CO_2$ and 3ppb for $CH_4$ in 5min. Remarkably, without any external calibration, the agreement in measured atmospheric absorbance spectra is better than $10^{-3}$, and the congruence of the retrieved concentrations is closed to the World Meteorological Organization (WMO) compatibility standards [9]. The reproducibility of open-path DCS is notably superior to its counterparts which typically exhibit uncertainties 10% or greater. Subsequently, all-fiber compact and portable dual-comb systems [10-12] have developed for fieldable application. Additionally, open-path mid-infrared DCS has been devised to probe the fundamental vibration modes of abundant gas-phase species with higher sensitive[13]. Recently, open-path DCS has been verified successfully in various application scenarios. These include detecting gas leaks in oil and gas production [14,15], monitoring volatile organic compounds in open air [16], measuring agricultural gas flux [17], estimating urban vehicle GHG emissions [18], and et al.

The effective path range of open-path DCS is constrained by its power detection ability. For sufficient power reception, a retroreflector, which folds the path, is generally necessary. Due to





collimated probing light and high-speed multi-heterodyne coherent detection, open-path DCS has achieved a remarkable range up to ten kilometers while maintaining portability and low-power usage [24,25], surpassing its counterparts. However, this range is still not sufficient for medium to large city scales. A bistatic configuration, which unfolds the path, could potentially double this range [40]. This requires synchronized time-base between transmitter and remote-receiver, adding complexities and reducing flexibility. While one hundred kilometers range has been demonstrated in such a configuration [26], the required watt-level comb light and the complexity of the system make it impractical for practical use, due to critical issues including safety concerns, cost, operability, and spectral bandwidth limitations. Extending the range to cover tens or even hundreds of kilometers for city scale emissions measurements remains a challenging yet desirable goal. And such extended ranges would also enable vertical column measurements from ground to stratospheric balloon or even satellite. To date, open-path DCS relies on the cooperation of a retroreflector or remote-receive as mentioned above, but the one fixed path limits the ability to dynamically measure regional trace gases distribution. Proposed solutions include multiple retroreflectors to create multiple paths [14,17] and aircraft-carried retroreflector for regional scanning [19]. However, open-path DCS still lacks the ability to freely choose sensing path by simply scanning the transmitted light orientation and utilizing backscatter signal from hard object, similar to lidar [20,21]. The realization of this non-cooperative targets open-path DCS is still hindered by current power detection ability. Given that open-path DCS transmits power typical in several to ten microwatts range, the resulting backscatter power from hard objects is substantially weaker compared to usual detection power of several milliwatts in standard DCS. An intuitive and rudimentary approach is to promote the transmitted power of comb light to several watt after amplifier [22,23], which is not eye-safety and has only been demonstrated for targets 20 meters away in lab setting. Besides, high-power comb light brings spectral non-linearity effects and high relative intensity noise and negatively impact high-precision spectroscopy detection. These issues also arise when pursuing longer range using high-power comb [26]. A pioneering study by Picqué et al. in 2020 [27] demonstrated the feasibility of photon-level dual-comb interference utilizing photon-counting techniques. Significantly, quantum interference persists across such a broad frequency range. Subsequently, only a few studies have explored photon-level DCS in weaker detection power [28]. Recent advancements [29], have expanded this approach to photon-level ultra-violate range





and visible range DCS, characterized by shot-noise-limited SNR and distinctly resolved comb lines. This approach substantially transcends limitations in sensitivity and SNR inherent in classical DCS to photon-level regime, providing an innovative way to achieve the ultra-long range or non-cooperative targets DCS open-path DCS.

In this photon-counting DCS framework, detected photon arrive-time events, or referred as photon clicks, encapsulate the statistical information pertinent to dual-comb interference. The reconstruction of an interferogram frame is achieved by statistically analyzing the time delays between the photon clicks and the designated start signal, which marks the commencement of each interferogram frame. As such, accurate determination of this precise start signal for each interferogram frame is critical for integrity and precision of the results. However, open-air turbulence fluctuations and fiber length wandering induces timing jitters in the arrival of interferograms, thereby impeding the reconstruction process from photon clicks. Mitigation of fiber length fluctuations can be partially achieved through careful coiling, secure affixing, and temperature control of the fibers. Nevertheless, jitter caused by turbulent atmosphere poses a more formidable challenge. Thus, the development of a novel setup, tailored for open-path sensing, alongside a trigger protocol capable of counteracting the effects of turbulence or fiber wandering, is imperative for effective application of the photon-counting method in open-path DCS. In this context, we propose a photon-level broadband dual-comb interferometer, for transcending classical detection limitations under turbulent open paths and noisy environments, facilitating the achievement of non-cooperative targets or ultra-long open-path DCS.

## Principle and Results
### concept of the open-air photon-level broadband Dual-comb interferometer

DCS utilizes two frequency combs with slightly different repetition frequencies to generate rapid, high-resolution spectral measurements through multi-heterodyne interference [30,32] (Fig. 1b,c). The interferogram could be straightly acquired from the electrical analog signal of a photon detector，enabling precise analysis across a broad spectrum without mechanical scanning [33]. Typically, DCS detection power is above 1 microwatt. In light-starving condition however, only several photons are detectable over extended periods, which need to be meticulously harnessed for interferogram reconstruction. As illustrated in Fig. 1a, after two comb lights are combined through a beam splitter, one output (referred to as "output 1" herein) is received by a photon-detector in





classical DCS detection condition. Each interferogram frame produced is then threshold-triggered at its center fringe (Fig. 1c). The timestamps from these triggers are fed into the start channel of a time-to-digital convert (TDC), initiating the computation of delay times relative to photon clicks from other channels. Another output (referred to as "output 2" herein) is directly employed to probe the atmosphere. The received dual-comb light, attenuated to photon-level either by reflection from hard targets or transmission through ultra-long path range, is spectrally dispersed and subsequently detected by an array of single-photon detectors (SPDs). This detection setup, which divides the task of broadband spectrum detection into several narrower, parallel segments, enhances the overall detection speed [34]. Photon clicks from each SPD are individually logged by the TDC. The arrival times of these clicks are statistically counted over the duration of a scan, a specific period initiated by the start signal. The accumulation of clicks continues until the interferogram is reconstructed with a desired SNR, after which it can be processed in a manner analogous to classical DCS.

This configuration, in conjunction with the photon-counting trigger protocol, has ability to facilitate in practical open-path sensing while effectively countering the time jitter between the start signal and photon clicks, a phenomenon induced by optical path length variation from fiber-length wandering or atmospheric turbulence (details in the **Methods**). The robustness of this photon-level broadband dual-comb interferometer against the time jitter is validated in the following.





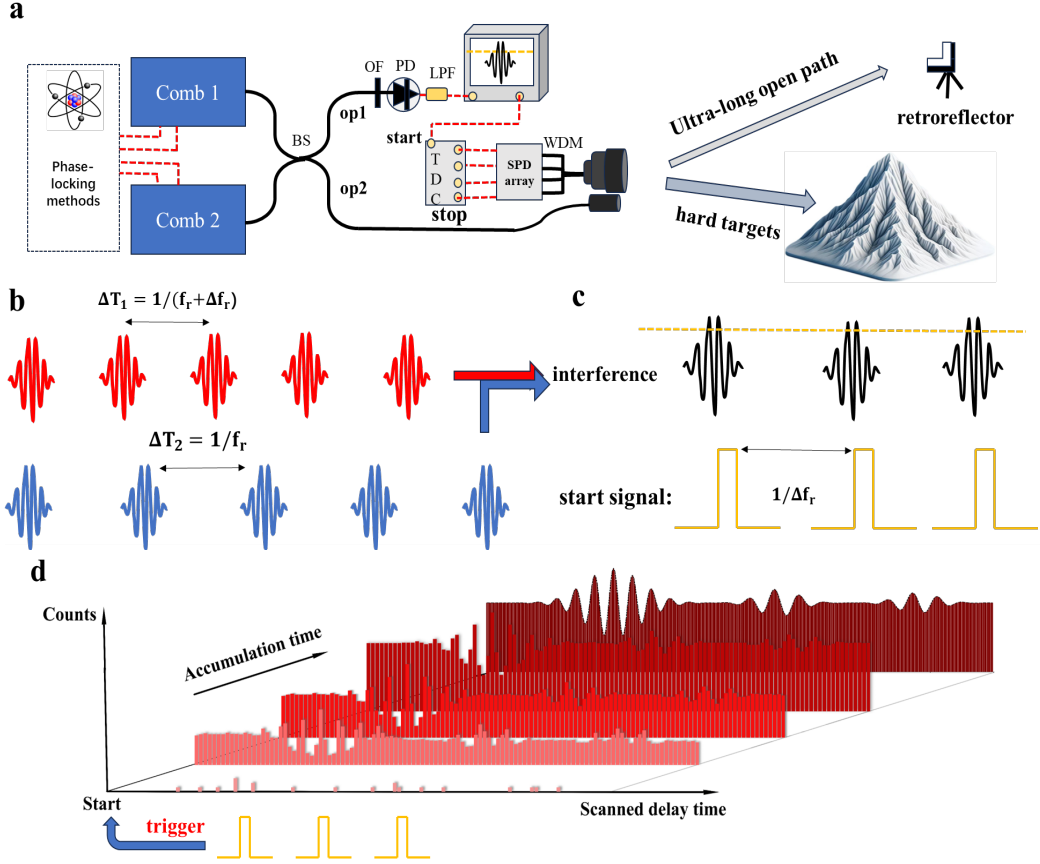

**Fig. 1 Photon-level broadband dual-comb interferometer senses the open-path trace gases. a** Schematic outline of the proposed system design. Two fiber mode-locked frequency combs are stabilized by phase-locking methods to stabilize the carrier-envelope frequencies $f_{ceo}$ and repetition frequency $f_r$. There exist a slightly repetition frequency difference $\Delta f_r$ between the two combs. Two combs light are combined through **BS** and divided into two outputs, with one for a classical DCS detection to produce the start signal and another for open-air photon-level sensing. Details see in the text. **BS**, beam splitter; **OP1/2**, output 1/2; **OF**, optical filter; **TDC**, time-to-digital convert; **PD**, photon detector; **WDM**, Wavelength Division Multiplexers; **LPF**, low-pass filter. **b** The principle of classical dual-comb interference in the time domain. The interval between laser pulses is dependent on comb's repetition frequency. **c** After the detection of interference process, the stream of interferogram frames is obtained, with a period of $1/\Delta f_r$. The start signal is produced by threshold-triggering center fringe of the interferogram. **d** Schematic of the photon-counting and the clicks accumulation process to reconstructed the photon-level interferogram. Unlike in classical DCS in adequate detection power, only a few photon clicks are detectable in one scan in the beginning. However, the arrival time of these clicks carry the statistical information of the dual-comb interference. Thus, the interferogram can be reconstructed after sufficient clicks in a proper statistical photon counting.

**Femtowatt level broadband DCS under conditions of severe optical path fluctuations**

The proposed photon-level broadband Dual-comb interferometer is applicable across various dual-comb systems and SPDs. In our experimental demonstration, two polarization-maintaining mode-lock fiber combs with a InGaAs single-photon detector, which are robustness, compactness, and field adaptability, are employed. We illustrate that even under highly light-starved condition,





the broadband HCN absorption spectroscopy information can be precisely extracted from photon-counting under unstable optical paths.

The carrier-envelope offset frequency $f_{ceo}$ of each comb is stabilized through self-reference. The two combs are tightly locked to a narrow-linewidth continues-wave (CW) laser at 193.4Thz as optical reference. The repetition frequencies for the two combs are 200.008Mhz and 200.001Mhz respectively, resulting in a 7.1Khz repetition frequency difference $\Delta f_r$. Each comb emits 4mW average power, center at 1550nm, with a broadband spectrum spanning 100nm. The emitted lights from both combs are combined directly in a fiber beam-splitter. One output of the beam-splitter is as output 1 to produce the start signals. The time interval between these start signals is $1/\Delta f_r$, which is about $141\mu s$ in the experiment. As shown in fig. 2b, the second output used as output 2, attenuated to less than 1nW after an attenuator, passes through a hydrogen cyanide standard gas pool ($H^{13}C^{14}N$ in SRM 2519a,16.5cm path length) and then through a free-space optical link. To simulate the turbulence and instability of an open-air path, the mirror, serving as the retroreflector is moving dynamically and continuously throughout the detection period. After receiving the reflected light from the mirror, the ideal setup requires several optical filters with different bandpass and SPDs to demonstrate the capability of piecewise spectrum parallel detection. Unfortunately, due to the lack of these hardware device, only a 10 nm broadband tunable optical filter and an InGaAs SPD are adopted as alternatives. By tunning the center bandpass of the filter, two segments of the HCN spectrum are obtained successively in two sets of detections. Ultimately, the SPD detects an average power of less than 20fW. Photon counting clicks from the SPD are recorded as the arrival times by another channel of the TDC. These clicks are statistically accumulated based on their delay time relative to the start signals. The digital resolution of the TDC for recording timestamps is set at 500 ps. Therefore, a simple digital low-pass filter can be applied to the statistical results to obtain the interferogram, functioning similarly to the low-pass electronic filter used in classical DCS (Fig. 2a, b). The dual-comb interferogram, reconstructed via photon counting statistics (hereafter referred to as "photon-level interferogram"), gradually emerges with increasing accumulation time.





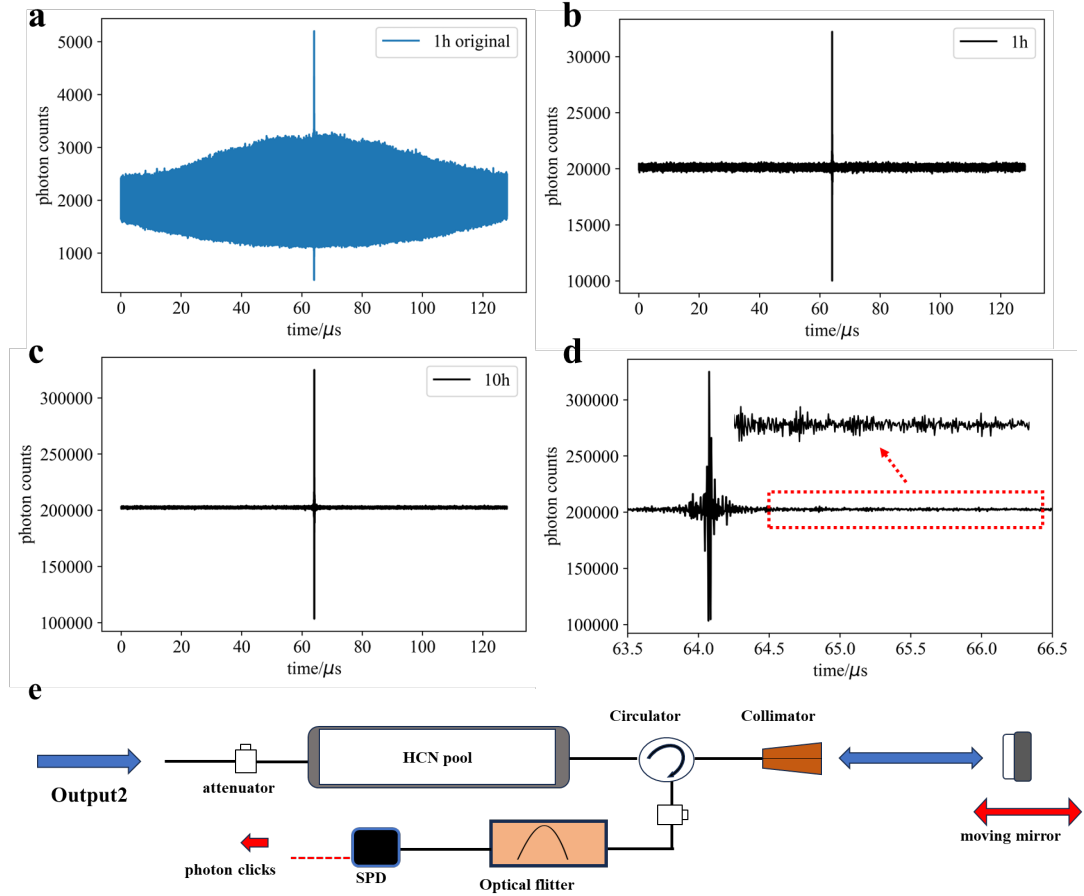

**Fig. 2 photon-level interferogram and experiment setup of the broadband HCN absorption spectroscopy. a** Original photon-counting accumulation results after 1h accumulation, covering a wavelength range from1549 nm to 1559 nm. Photon clicks are counted using a 500ps digital resolution. **b.** The photon-level interferogram results are derived by post-processing with a 100Mhz digital low-pass filter. **c** The photon-level interferogram after 10h accumulation. **d** serves as an enlargement of **c.** as accumulation, time-domain information on gas molecule absorption features progressively emerge behind the interferogram's central fringe. **e** Experimental setup following the output 2, designed to emulate gas absorption and drastic changes in optical path length typical of open-air environments. Two comb lights pass through an attenuator, a 16.5cm $H^{13}C^{14}N$ standard gas pool, a circulator and a collimator, then travel to free-space, reflected back by a mirror. The mirror, mounted on a motorized stage, moves forward and backward continuously, completing a 48 mm round trip at a speed of 4 mm/s. The reflected light, after repassing the circulator and attenuator and attenuated to femtowatt levels, is detected by an InGaAs single photon detector (SPD). Photon clicks from the SPD reconstruct the interferogram.

In the first phase of detection, the bandpass of the tunable filter is set to span from 1548-1558 nm, featuring a flat-top shape. A photon-level interferogram after 10 hours of accumulation is shown in Fig. 2c. The counting rate for the SPD stand at approximately 147.5 k/s. Despite the ultra-low detection rate—averaging one click per thousand laser pulses, the multi-heterodyne information, encapsulating the beating between thousands of different frequencies, was successfully extracted from the photon-counting clicks. According to 35.5% quantum efficiency





of the SPD, the average power from each comb before SPD is calculated to be 53 fW, and the average power per comb line is estimated to be about 4.3 aW. This is a remarkable $10^{10}$ times weaker than the power typically employed in classical DCS. Fourier transform of the photon-level interferogram reveals distinctive hydrogen cyanide absorption features. The intensity spectrum displayed 13 narrow and deep absorption features, corresponding to the 2v3 band of $H^{13}C^{14}N$, specifically the P9-P21 absorption lines in P branch (Fig. 3a, right-side intensity spectrum).

In the second phase of detection, the 10nm bandpass center is shifted to 1533nm, covering a range from 1528 nm to 1538 nm. It is important to note that this range falls within the adjacent Nyquist sampling range of the first detection phase. To ensure coherence between the start signals and the photon-counting clicks, the narrow bandpass center in output 1 channel is adjusted to 1535 nm. The counting rate slightly decrease to about 143.5 k/s, which is attributed to a lower comb spectrum intensity within this bandpass range compared to the first detection phase. The power for each comb before the detector is calculated to be approximately 52 fW and the average power per comb line is estimated to be around 4 aW. The intensity spectrum of this phase detection predominantly covers the 2v3 band of $H^{13}C^{14}N$, R6-R24 absorption lines in the R branch, although R20-R24 lines are less discernible amidst baseline noises (Fig. 3a, left-side intensity spectrum, after 10h accumulation). Several baseline removal algorithms such as polynomial fitting [7,9], m-Fid cepstral analysis [35] and machine learning [36], could be employed the transmittance spectrum to extract the absorbance spectrum, which correlates linearly with concentrations and facilitates gases parameters retrieval. Here, a simple piecewise polynomial fitting method was applied to the intensity spectrum, or the transmittance spectrum more specifically, after 32h accumulation. The total 20nm broadband photon-level absorbance spectrum, encompassing both P- and R- branches in two piecewise segments, along with its comparison with Hitran2024 database, is presented in Fig 3b. The discrepancy between detected and calculated theoretical spectrum manifests as dominated stochastic noises with a standard deviation that is closed to 10% of the average absorbance (Fig. 3c). This deviation is primarily attributed to the ultra-low counting rate, resulting in a relatively low SNR. Additionally, factors like temperature fluctuations, shifts in the comb spectrum and optical reference over long-period detection, as well as uncontrolled ambient noise from background light and laboratory instrument vibrations, also contribute to this variation.





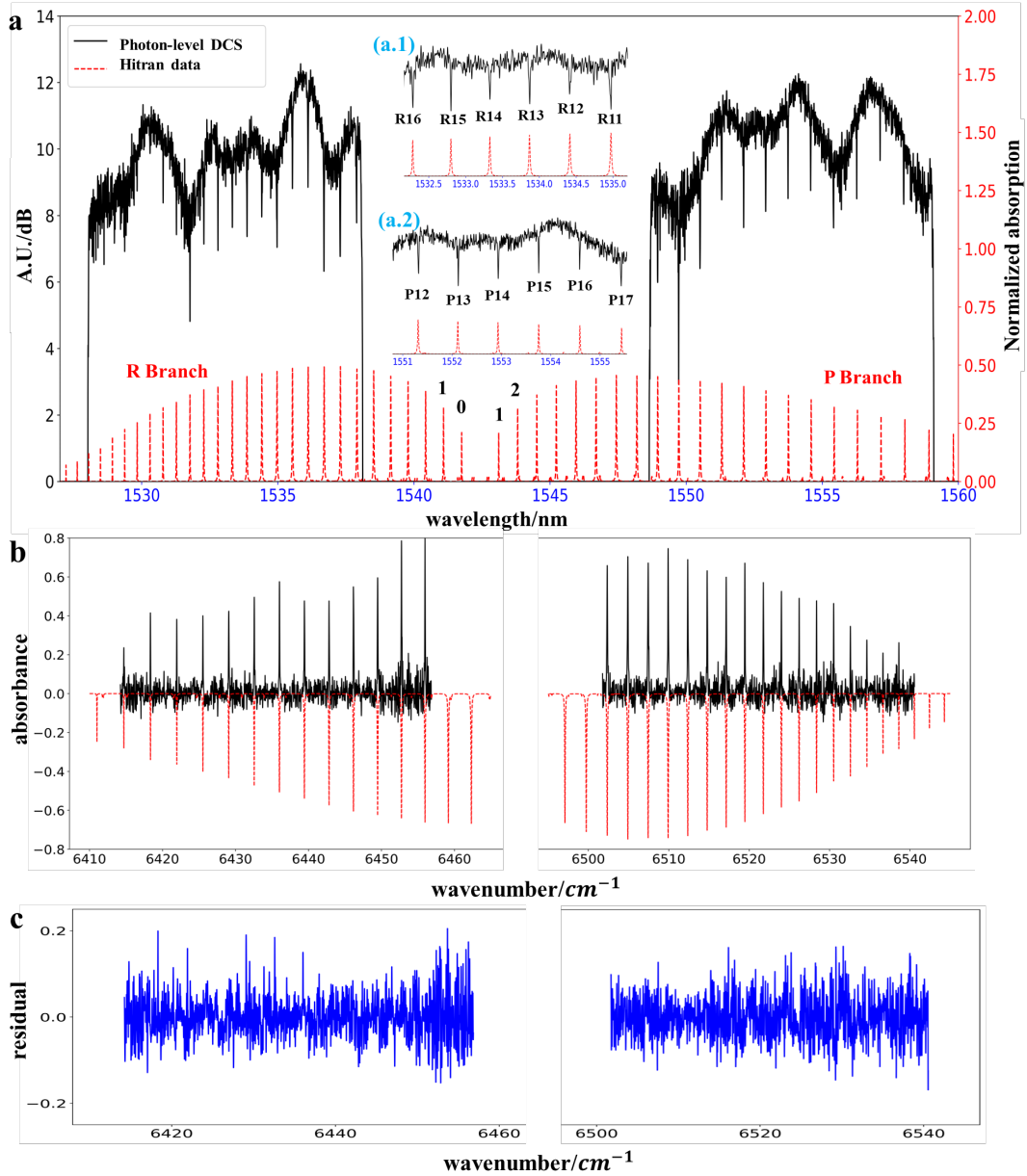

**Fig. 3 Photon-level broadband dual-comb absorption spectrum of H¹³C¹⁴N under optical path jittering condition. a** The 20 nm broadband photon-level HCN transmittance spectrum is presented in log scale as black solid lines, alongside a comparison with the theoretical absorbance spectrum depicted as red dashed lines. The spectrum is acquired from two sets of detections in different range (1528-1538nm and 1548-1558nm), with 10h accumulation each. This spectrum encompasses the R- and P- branches of the 2v3 band of H¹³C¹⁴N. The detection power is less than 20 fW. Insets (a.1) and (a.2) provide detailed zoomed-in plots of the R11-R16 and P12-P17 regions. **b** After baseline removal from the transmittance spectrum accumulated over 32 hours, the photon-level absorbance spectrum is obtained, shown as black solid lines. The H¹³C¹⁴N gas pool in the experiment conforms to SRM 2519a standard and features a 16.5cm optical path. Thus, thermotical absorbance spectrum is calculated based on the Hitran2024 model (represented as red dashed lines, inverted for clarity. Also represented as red dashed lines in a). **c** The residual between detected and calculated absorbance spectrum in **b**.

## Evaluating comb-line resolved photon-level DCS with shot-noise-limit SNR





By selecting the start signals at every 48th trigger count, a desired photon-level interferograms with a $7500\mu s$ scanning time, incorporating 48 center fringes, is reconstructed (Fig. 4a). The Radio Frequency (RF) spectrum, directly derived from the Fourier transform of this interferogram, distinctly exhibits comb lines at intervals of $f_r$ (Fig. 4b, c, d). The line-width of these comb lines is approximately 140 Hz, aligning with the Fourier transform limit of the interferogram. This underscores the high coherence maintained between photons from the two combs over the scanning period.

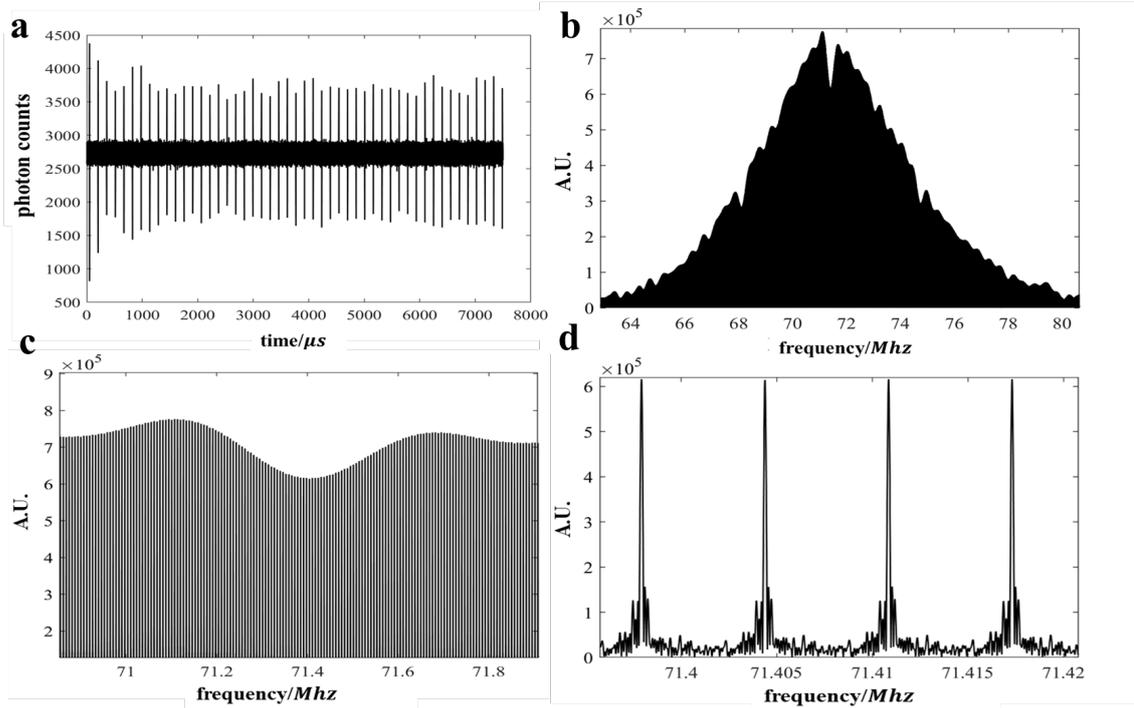

**Fig. 4 comb-line resolved photon-level DCS.** The experiment conditions are consistent with those in Fig. 3, except for two changes: the optical filter before the SPD is replaced with a Gaussian shape filter centered at 1551.3nm with 1.8nm Full Width at Half Maximum (FWHM), and the repetition frequency difference $\Delta f_r$ is 6.5khz. **a** The photon-level interferogram, including 48 center fringes, is obtained after 7h accumulation, with each scan lasting $7500\mu s$. The counting rate of SPD is about 162k/s. Variations in the height of the interferogram's center fringes can be attributed to the feedback-induced relaxation in the phase-locking process. **b** The Radio Frequency (RF) spectrum is derived by performing Fourier transform on the photon-level interferogram. **c** A detailed view of the central RF spectrum from **b**. the height variation of spaced comb-lines highlights the absorption feature from 2v3 P12 line. **d** The RF spectrum clearly exhibits equally spaced comb lines with an interval of $\Delta f_r$. The line-width of each comb line is 140Hz, approaching the Fourier transform limitation.

The use of a cavity-stabilized, narrow linewidth CW laser as the optical reference significantly enhances the mutual coherence of the dual-comb source. This ensures the achievement of comb-line resolved photon-level DCS after long-period accumulation. In comparison, the photon-level DCS is obtained with the two combs' repetition frequencies straightly locked to the Rb atomic





clocks as the reference. In this phase-locking setup, the relative linewidth between the two combs exceeds the $\Delta f_r$. The resulting photon-level interferogram reveals that only the initial center burst fringe is present, while subsequent center-burst fringes are neutralized over the course of the accumulation (Supplementary Fig. 2). In the frequency domain, the overlapping comb lines preclude the resolution of individual comb lines. Although the HCN absorption features can still be discriminated in the photon-level spectrum, the resolution is notably reduced. It is worth notice that in this start signal triggering protocol, $\Delta f_{ceo}$ and $\frac{f_r}{\Delta f_r}$ are not required to be zero and an integer, respectively (Supplementary Fig. 1). This relaxation in setup constraints simplifies the requirements for the locking and electronic synchronization systems, enhancing compatibility with various types of dual-comb systems.

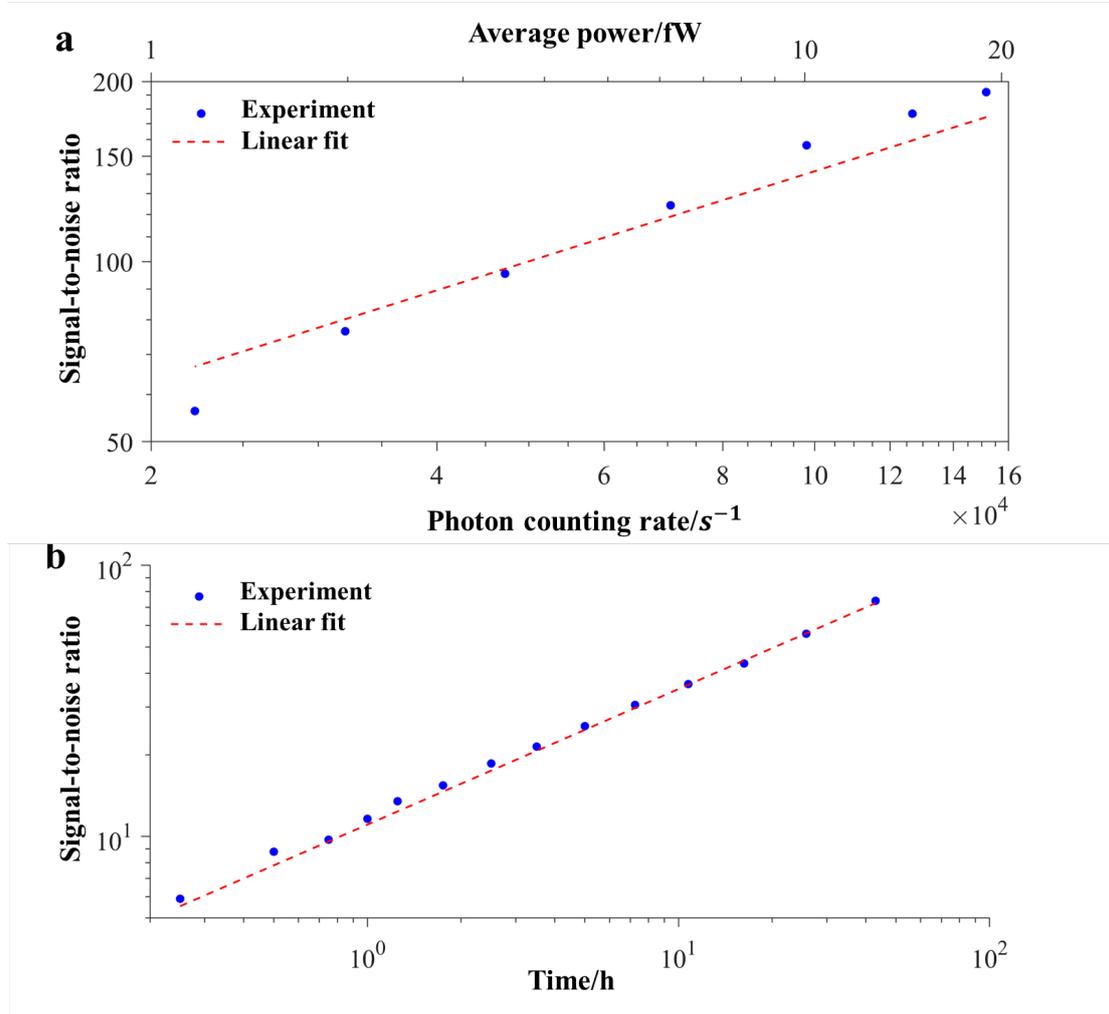

**Fig. 5 Long-period coherent adding stability and quantum-noise-limited SNR of the interferogram. a** The SNR scales with the square-root of the photon counting rate, which varies from 1 to 20 fW equivalently. For each set of detection in different photon counting rate, the accumulation time is held constant at 4h. The scaling trend's slope is slightly above 0.5, as indicated by the red dotted line with a slope of 0.5, attributed to changes in single-





photon detector's working conditions over the extended test period, such as dark counting rate and SPD bias voltage. **b** The SNR is proportional to the square root of the accumulation time, with a photon counting rate maintained at 105.2 k/s (equivalent to 13.5 fW received by the SPD, and 38 fW before the SPD in consider of 35.5% quantum efficiency). The red dotted line represents the linear fit, characterized by a slope of 0.5.

The SNR of the photon-level interferogram shows a direct proportionality to square roots of the accumulation time, exhibiting a proportional constant of 0.5. This coherent addition pattern is observed to be sustainable for up to 60 hours under optical path jitter conditions, with the limitation being our acquisition duration (Fig. 5b). This trend of coherent addition is expected to continue with prolonged accumulation times. Further, the SNR of the photon-level interferogram scales proportionally with the square root of SPD's counting rate, consistent with the regularities of shot-noise limit detection (Fig. 5a). And the experimentally calculated SNR aligns closely with the theoretical SNR (details in table. 1 and the **Method**).

| | M | V | Scan time $T_{scan}$ ($\mu s$) | L | $T_{total}$ | Experimental SNR | Theoretical SNR | Average Power per comb line before SPD (aW = $1 \times 10^{-18}\,W$) |
|---|---|---|---|---|---|---|---|---|
| Fig 3a (P branch) | 6219 | 0.60 | 128 | 1 | 10 | 7.63 | 7.88 | 4.3 |
| Fig 3a (R branch) | 6382 | 0.56 | 128 | 1 | 10 | 6.90 | 7.02 | 4 |
| Fig 3b (P branch) | 6219 | 0.60 | 128 | 1 | 32 | 13.99 | 14.01 | 4.3 |
| Fig 3b (R branch) | 6382 | 0.56 | 128 | 1 | 32 | 12.15 | 12.69 | 4 |
| Fig 4 | 2200 | 0.44 | 7500 | 48 | 7 | 14.47 | 15.08 | 18.5 |

**Table 1. Experimentally measured and theoretical calculated signal-to-noise ratio**. Discrepancies between the theoretical and experimental SNR values can be attributed to additive noises present in the experimental setup and variations in the intensity of comb lines, which are not accounted for in the theoretical model. The average power per comb line is determined by halving the total detected power (derived from the photon counting rate and the SPD's quantum efficiency of 35.5%) and then dividing this value by the number of comb lines.

## Discussion and Conclusions

We have demonstrated broadband dual-comb spectroscopy spanning 10nm at an average detected power less than 20fW, with the power per comb line reaching as low as 4 aW. Under such extreme light-starved conditions, the accumulation time correspondingly increases, underscoring the importance of long-period coherent stability in the photon-level dual-comb interferogram, particularly when dealing with continuously varying optical paths. Our results show that the proposed dual-comb interferogram can reconstruct high-resolution, high-precision





spectra, even under these challenging detection and environmental conditions. The experiment utilized a single SPD, resulting in a relatively long acquisition time. However, this time can be substantially reduced by employing multiple single-photon detectors for parallel detection [34]. Besides, this parallel detection setup allows for implementing a higher $\Delta f_r$, which could further accelerate the accumulation speed. The counting rate in our study was kept below 150 k/s to verify the long-period stability of the proposed interferometer. In practical applications, however, the counting rate could be significantly increased to shorten the acquisition time, subject to the saturation counting rate of SPD. Advances in III–V compound semiconductor-based avalanche photodiodes and superconducting nanowire detectors, are anticipated to enable SPDs with rapid photon counting, ultra-low dark counts, compact size, low cost, and ease of operation [38,39]. In classical DCS, SNR is constrained by relative intensity noises (RIN), detector dynamic range or detector noise, which are far above shot-noise limitation and hard to breakthrough [34]. Here, the SNR in our photon-level DCS achieves shot-noise-limit, which scales to the square roots of the counting rate of SPD.

    In traditional open-path DCS, lasers from two combs in a common-mode transmission through the atmosphere effectively suppresses phase noises induced by turbulent air. In photon-level dual-comb interferometer, this common-mode configuration not only mitigates phase noises from the subsequent optical path, such as fiber-length wandering and open-air turbulence, but also synchronizes the start signal with photon clicks in photon-counting process. This synchronization is crucial for successfully reconstructing the photon-level interferogram. In this configuration, the depth of gas absorption is doubled compared to that in a single comb laser probing setup, thereby enhancing the sensitivity of gas detection. The single comb laser probing setup detects both absorption and dispersion responses of gases [7,8]. In contrast, our configuration only captures the absorption spectrum. However, it is worth noting that the dispersion spectrum can be inferred from the absorption spectrum by employing the Kramers-Kronig relationship [37,8]. In classical DCS, a tightly locked dual-comb interferometer exhibits high mutual coherence between combs, allowing for the instrument's line shape to be ignored, and enabling the direct acquisition of the spectroscopy's absolute frequency without the need for calibration [41,42]. This is pivotal for distinguishing multiple trace gases and determining their concertation, given that trace gas absorption lines are characterized by specific central frequencies and narrow line shapes. In





photon-counting DCS, maintaining high mutual coherence is equally crucial for the accurate extraction of spectral information from photon clicks, ensuring both high spectral resolution and signal-to-noise ratio. In our study, the optical reference employed is a narrow line-width CW laser stabilized to a vacuum cavity, requiring a controlled laboratory environment. Looking ahead, the cavity-stabled laser could be replaced with a free-running diode laser based on a "bootstrapped" frequency referencing scheme [11]. Such a free-running diode laser-based dual-comb interferometer would offer robustness and compactness for field applications, and its performance is demonstrated to be comparable to that of a cavity-stabilized laser reference in laboratory settings.

In conclusion, we have proposed a photon-level dual-comb interferometer employing a common-mode configuration, capable of achieving highly precise photon-level Dual-Comb Spectroscopy (DCS) in turbulent open-air conditions, while also being field-deployable with compact all-fiber components. Utilizing this interferometer, we have demonstrated photon-level broadband dual-comb spectroscopy under challenging environmental conditions, achieving the lowest detection power and the broadest spectral bandwidth recorded to date, and marking a first in the infrared range detection. Our results establish that shot-noise-limited and comb-line-resolved DCS is feasible, even under adverse conditions characterized by dynamically changing open path lengths and uncontrolled ambient interferences such as vibrations and background light. This interferometer which combines miniaturization, portability, low-power consumption and eye safety, paves the way to qualify trace gases in turbulent open-air in various appealing sensing forms, such as non-cooperative targets sensing, hundreds of kilometers sensing range, and open-path link from ground to the satellite.

**Methods**

**photon-level dual-comb interferometer system**

Two home-made erbium-doped-fiber mode-lock femtosecond lasers with about 200Mhz repetition frequency are used as dual-comb light source. Each laser emits about 4mW light power and the spectrum spans from 1500-1600 nm. A homemade ultra-stable laser stabilized to a vacuum cavity at 193.39Thz is used as the optical reference for dual comb system. The frequency signals of the dual-comb interferometer phase-locking system are all referred to a Rb atomic clock





(SRS, FS725), The carrier-envelope offset frequency $f_{ceo} = 250Mhz$ of each comb is stabilized by using f-2f self-referencing method. And two comb lines, from two combs each, located on two sides of the ultra-stable laser, are phase-locked to the ultra-stable laser at $f_{beat} = 270Mhz$. The repetition frequencies of two combs are about 200Mhz with about 7.1khz difference.

In photon-level DCS experiment, the gas pool (WaveLength, HCN-13-25) is a fiber coupled gas cell with 16.5cm path length in NIST-traceable standard 2519a. One output of the beam-splitter is attenuated to about $10\mu W$ average power and subsequently detected by a diode photon-detector as the start signal channel. A narrow fiber coupled tunable optical filter (WL Photonics) with 1.8nm FWHM Gaussian shape is used before one of the pin-diode in balanced photon detector (Thorlabs PDB570C) to satisfy Nyquist sampling condition and also acquire the interferogram frames in a relative high SNR. An electronic low-bandpass filter after the photon-detector is used to ensure the one-to-one comb-line multi-heterodyne. In photon-level detection channel, a programmable optical filter (Waveshaper 1000A) is used before the InGaAs SPD (QuantumCTek, QCD600B-H, quantum efficiency is 35.5%). The TDC (Time Tagger Ultra) is used for calculating the delay time between trigger signal and photon clicks. In this system, the carrier-envelope offset frequency difference $\Delta f_{ceo}$ need not to be zero (Supplementary Fig. 1). The photon-level dual-comb interferometer system can be phase-locked without the optical reference by locking the repetition frequency straight at 200Mhz, referred to the Rb atomic clock. However, this degrades the relative coherence of two combs, and the spectral resolution of photon-level DCS becomes lower than tightly locking to the optical reference (Supplementary Fig. 3).

**Theoretical shot-noise-limited SNR in photon-counting statistic in different scanning length**

According to ref. [29], the SNR of the photon-level dual-comb interferogram at zero optical delay (t=0) constrained by quantum-noise-limited is:

$$\left(\frac{S}{N}\right)_{t=0} = \frac{n_{\text{interf}}}{\sqrt{n+n_{\text{interf}}}} \tag{1}$$

Where $(n + n_{\text{interf}})$ represents the total number of photons counts at t=0 bin and the shot noise in statistic is $\sqrt{n+n_{\text{interf}}}$ . $n_{\text{interf}}$ is the number of photons counts from the maximum aptitude of the interference signal. In other words, the maximum and minimum photon counts are $n_{\text{interf, max}} = n + n_{\text{interf}}$ and $n_{\text{interf, min}} = n - n_{\text{interf}}$ respectively. And the interference visibility V is





$$V = \frac{n_{\text{interf, max}} - n_{\text{interf, min}}}{n_{\text{interf, max}} + n_{\text{interf, min}}} = \frac{n_{\text{interf}}}{n} \qquad (2)$$

The SNR in the frequency domain at a specific frequency $\nu$ is related to $\left(\frac{S}{N}\right)_{t=0}$ and can be

deduced as:

$$\left(\frac{S}{N}\right)_{\nu} = \sqrt{\frac{2}{K_s} \frac{B(\nu)}{\bar{B}_e}} \left(\frac{S}{N}\right)_{t=0}$$

$$= \sqrt{\frac{2}{K_s}} \cdot \frac{\frac{1}{M}}{\frac{1}{K_s}} \cdot \frac{\text{V}}{\sqrt{1+V}} \cdot \sqrt{n} \qquad (3)$$

Where M is the number of comb lines and $K_s$ is the total scanned bins. Assuming the number of

bins for a frame of interferogram is $K$.

*When $K_s < K$ (only one interference center fringe of the photon-level interferogram is scanned):*

$$\left(\frac{S}{N}\right)_{\nu} = \sqrt{2} \cdot \frac{\sqrt{K_s}}{M} \cdot \frac{\text{V}}{\sqrt{1+V}} \cdot \sqrt{N_{CR} \cdot \frac{T_{total}}{K}}$$

$$= \sqrt{2} \cdot \frac{1}{\text{M}} \cdot \frac{\text{V}}{\sqrt{1+V}} \cdot \sqrt{N_{CR} \cdot T_{eff}} \qquad (4)$$

Where $N_{CR}$ is the counting rate, $T_{total}$ is total accumulation time and $T_{eff} = \frac{K_s}{K} \cdot T_{total}$ is the

effective accumulation time for the scanned photon-level interferogram.

*When $K_s > K$ (the photon-level interferogram with L period fringes is scanned):*

$$\left(\frac{S}{N}\right)_{\nu} = \sqrt{2} \cdot \frac{\sqrt{K_s}}{M} \cdot \frac{\text{V}}{\sqrt{1+V}} \cdot \sqrt{N_{CR} \cdot \frac{T_{total}}{L \cdot K}}$$

$$= \sqrt{2} \cdot \frac{1}{\text{M}} \cdot \frac{\text{V}}{\sqrt{1+V}} \cdot \sqrt{N_{CR} \cdot T_{eff}} \qquad (5)$$

Where $T_{eff} = \frac{K_s}{L \cdot K} \cdot T_{total}$.

**The start signal jitter in different configurations**

There exist another two interferometer configurations that use the interferogram-triggered start

signal protocol (Supplementary Fig. 4). One setup (referred as "setup 1 hereafter) is that one comb

pass through a gas pool firstly as the signal comb, and then combines with another local comb in a

beam splitter, as demonstrated in references [27,29]. In setup 1, one output is as the start signal

channel. And another output is as photon-level DCS detection channel, which needs to be

attenuated to photon level firstly. In other words, the photon-level DCS detection channel could





have been treated as a classical DCS. This setup makes no sense in photon-level open-path DCS sensing, although it is proved by experiment that the time jitter between the start signal and photon clicks is the smallest (Supplementary Fig. 5a,b,c). Second setup (referred as "setup 2" hereafter) is similar as the phase-sensitive open-air DCS. The start signal channel is an individual part by beating the two combs directly. However, it is proved that the time jitter between the start signal and photon clicks due to the fiber wandering is high enough to erase the dual-comb multi-heterodyne information in the photon counting process, not to mention the impact of turbulent atmosphere (Supplementary Fig. 5e). By using the proposed common-mode setup here (referred as "setup 3" hereafter), the time jitter can be controlled within a lower level near the level of setup 1 in even the optical path-length jitter condition.

## Data availability

All data supporting the main findings of this work are available within the paper and its Supplementary Information, or available from the corresponding author upon reasonable request.

## References

1.  Waxman, Eleanor M, Finneran, et al. "Gas-phase broadband spectroscopy using active sources: progress, status, and applications[J]". Journal of the Optical Society of America, B. Optical Physics, 2017.

2.  Lin C H, Johnston C T, Grant R H, et al. "Application of Open Path Fourier Transform Infrared Spectroscopy (OP-FTIR) to Measure Greenhouse Gas Concentrations from Agricultural Soils"[J]. 2018.DOI:10.5194/amt-2018-373.

3.  T. L. Marshall, C. T. Chaffin, R. M. Hammaker, and W. G. Fateley, "An introduction to open-path FT-IR atmospheric monitoring," Environ. Sci. Technol. 28, 224A–232A (1994).

4.  U. Platt and J. Stutz, "Differential Absorption Spectroscopy", Physics of Earth and Space Environments (Springer, 2008).

5.  U. Platt and D. Perner, "Measurements of atmospheric trace gases by long path differential UV/visible absorption spectroscopy," in Optical and Laser Remote Sensing, D. D. K. Killinger and D. A. Mooradian, eds., Vol. 39 in Springer Series in Optical Sciences (Springer, 1983), pp. 97–105.

6.  Bo Yu, Xu Wu, Minghui Zhang, Tianbo He, Jingsong Li, "Tunable diode laser absorption spectroscopy for open-path monitoring gas markers in fire combustion products", Infrared Physics & Technology, Volume 131, 2023,104690, ISSN 1350-4495.

7.  G. B. Rieker, F. R. Giorgetta, W. C. Swann, J. Kofler, A. M. Zolot, L. C. Sinclair, E. Baumann, C. Cromer, G. Petron, C. Sweeney, P. P. Tans, I. Coddington, and N. R. Newbury, "Frequency-comb-based remote sensing of greenhouse gases over kilometer air paths," Optica 1, 290-298 (2014)






8.  Fabrizio, et al. "Broadband Phase Spectroscopy over Turbulent Air Paths." Physical Review Letters 115.10(2015):103901-103901.

9.  Waxman E M, Cossel K C, Truong G W, et al. "Intercomparison of open-path trace gas measurements with two dual-frequency-comb spectrometers"[J]. Atmospheric Measurement Techniques, 2017, 10(9):3295-3311. DOI:10.5194/amt-10-3295-2017

10. L. C. Sinclair, J.-D. Deschênes, "Invited Article: A compact optically coherent fiber frequency comb", Rev. Sci. Instrum. 86, 081301 (2015).

11. Gar-Wing, et al. "Accurate frequency referencing for fieldable dual-comb spectroscopy." Optics Express (2016).

12. L. C. Sinclair, I. Coddington, W. C. Swann, G. B. Rieker, A. Hati, K. Iwakuni, and N. R. Newbury, "Operation of an optically coherent frequency comb outside the metrology lab," Opt. Express 22, 6996-7006 (2014).

13. Ycas, G., Giorgetta, F.R., Baumann, E. et al. "High-coherence mid-infrared dual-comb spectroscopy spanning 2.6 to 5.2 μm". Nature Photon 12, 202–208 (2018).

14. Sean Coburn, Caroline B. Alden, Robert Wright, Kevin Cossel, Esther Baumann, Gar-Wing Truong, Fabrizio Giorgetta, Colm Sweeney, Nathan R. Newbury, Kuldeep Prasad, Ian Coddington, and Gregory B. Rieker, "Regional trace-gas source attribution using a field-deployed dual frequency comb spectrometer," Optica 5, 320-327 (2018).

15. Alden, Caroline B , et al. "Single-Blind Quantification of Natural Gas Leaks from 1 km Distance Using Frequency Combs." Environmental Science & Technology 53.5(2019):2908-2917.

16. Gabriel Ycas, Fabrizio R. Giorgetta, Kevin C. Cossel, Eleanor M. Waxman, Esther Baumann, Nathan R. Newbury, and Ian Coddington, "Mid-infrared dual-comb spectroscopy of volatile organic compounds across long open-air paths," Optica 6, 165-168 (2019)

17. Herman, Daniel I. , et al. "Precise multispecies agricultural gas flux determined using broadband open-path dual-comb spectroscopy." Science Advances 14(2021)

18. Waxman, Eleanor M., et al. "Estimating vehicle carbon dioxide emissions from Boulder, Colorado using horizontal path-integrated column measurements." Atmospheric Chemistry and Physics (2018):1-23.

19. Kevin C. Cossel, Eleanor M. Waxman, Fabrizio R. Giorgetta, Michael Cermak, Ian R. Coddington, Daniel Hesselius, Shalom Ruben, William C. Swann, Gar-Wing Truong, Gregory B. Rieker, and Nathan R. Newbury, "Open-path dual-comb spectroscopy to an airborne retroreflector," Optica 4, 724-728 (2017).

20. Mcmanamon, Paul F. . LiDAR Technologies and Systems. 2019.

21. Fujii, Takashi , and T. Fukuchi . "Laser remote sensing." Optical Engineering volume 21.116(2005):415-417(3).

22. Martin Godbout, Jean-Daniel Deschênes, and Jérôme Genest, "Spectrally resolved laser ranging with frequency combs," Opt. Express 18, 15981-15989 (2010)

23. Sylvain Boudreau, Simon Levasseur, Carlos Perilla, Simon Roy, and Jérôme Genest, "Chemical detection with hyperspectral lidar using dual frequency combs," Opt. Express 21, 7411-7418 (2013)

24. G. . -W. Truong et al., "Dual-comb spectroscopy for city-scale open path greenhouse gas monitoring," *2016 Conference on Lasers and Electro-Optics (CLEO)*, San Jose, CA, USA, 2016, pp. 1-2.

25. K. C. Cossel, F. R. Giorgetta, E. Baumann, W. C. Swann, I. Coddington, and N. R. Newbury, "28 km Open Path Dual-Comb Spectroscopy," in *OSA Optical Sensors and Sensing Congress 2021 (AIS, FTS, HISE, SENSORS, ES)*, S. Buckley, F. Vanier, S. Shi, K. Walker, I. Coddington, S. Paine, K. Lok Chan, W. Moses, S. Qian, P. Pellegrino, F. Vollmer, G. , J. Jágerská, R. Menzies, L. Emmenegger, and J. Westberg, eds., OSA Technical Digest (Optica Publishing Group, 2021), paper EW3C.5.







26.   Jin-Jian Han, Wei Zhong, et al. "Dual-comb spectroscopy over 100km open-air path". arXiv:2310.19294

27.   Picqué N, Hänsch TW. "Photon-level broadband spectroscopy and interferometry with two frequency combs". Proc Natl Acad Sci U S A. 2020 Oct 27;117(43):26688-26691.

28.   Hu, H.; Ren, X.; Wen, Z.; Li, X.; Liang, Y.; Yan, M.; Wu, E. Single-Pixel Photon-Counting Imaging Based on Dual-Comb Interferometry. *Nanomaterials* 2021, *11*, 1379.

29.   Bingxin Xu, Zaijun Chen, et al."Near-ultraviolet photon-counting dual-comb spectroscopy". arXiv:2307.12869

30.   Coddington, W. C. Swann, and N. R. Newbury, "Coherent linear optical sampling at 15 bits of resolution," Opt. Lett. 34, 2153-2155 (2009)

31.   Coddington I., W. Swann, et al. "Coherent Dual Comb Spectroscopy at High Signal to Noise." Physical Review A 82.4(2010):3535-3537.

32.   Picqué, N., Hänsch, T.W. "Frequency comb spectroscopy". Nature Photon 13, 146–157 (2019).

33.   Coddington I, Newbury N, Swann W. "Dual-comb spectroscopy". Optica. 2016;3(4)

34.   Nathan R. Newbury, Ian Coddington, and William Swann, "Sensitivity of coherent dual-comb spectroscopy," Opt. Express 18, 7929-7945 (2010)

35.   Ryan K. Cole, Amanda S. Makowiecki, Nazanin Hoghooghi, and Gregory B. Rieker, "Baseline-free quantitative absorption spectroscopy based on cepstral analysis," Opt. Express 27, 37920-37939 (2019)

36.   Chen Xinyi, Huang Chao, et al. "Phase-Sensitive Open-Path Dual-Comb Spectroscopy with Free-Running Combs," Phys. Rev. Applied 19, 044016

37.   Thomas G. Mayerhöfer, Vladimir Ivanovski, Jürgen Popp," Infrared refraction spectroscopy - Kramers-Kronig analysis revisited", Spectrochimica Acta Part A: Molecular and Biomolecular Spectroscopy,Volume 270,2022,120799,ISSN 1386-1425.

38.   Zhang, J., Itzler, M., Zbinden, H. et al. Advances in InGaAs/InP single-photon detector systems for quantum communication. Light Sci Appl 4, e286 (2015).

39.   Hadfield, R.H. Superfast photon counting. Nat. Photonics 14, 201–202 (2020).

40.   F. R. Giorgetta, E. Baumann, B. R. Washburn, N. Malarich, J. -. Deschênes, I. Coddington, N. R. Newbury, and K. C. Cossel, "Dual-Comb Spectroscopy of Carbon Dioxide and Methane Across a 14.5 km Long Outdoor Path," in *Optica Sensing Congress 2023 (AIS, FTS, HISE, Sensors, ES)*, Technical Digest Series (Optica Publishing Group, 2023), paper FTh2A.2.

41.   Chen, Z., Yan, M., Hänsch, T.W. et al. A phase-stable dual-comb interferometer. Nat Commun 9, 3035 (2018). https://doi.org/10.1038/s41467-018-05509-6

42.   Zaijun Chen, Theodor W. Hänsch, Nathalie Picqué. "Mid-infrared feed-forward dual-comb spectroscopy". Proc Natl Acad Sci U S A. 2019 Feb 12;  116 (9) 3454-3459.


## Acknowledgments


This work was supported by the National Natural Science Foundation of China (42125402,42188101), Innovation Program for Quantum Science and Technology (2021ZD0300300), the Project of Stable Support for Youth Team in Basic Research Field, CAS (YSBR-018), Anhui Initiative in Quantum Information Technologies(AHY140000). We thank Dr. Li zhengping (University of Science and Technology of China) for the discussion of diffusion








## Competing interests

The authors declare no competing interests.

\





## Supplementary figures

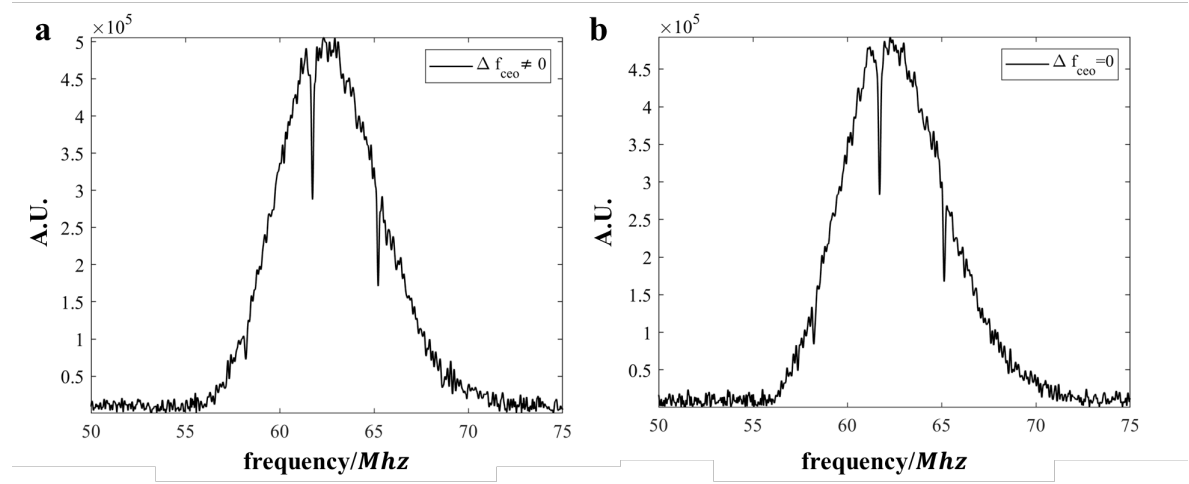

**Fig. 1** The photon-level Dual-Comb Spectroscopy (DCS) is achievable regardless of whether the carrier-envelope offset frequency difference (Δf_ceo) is zero. **a** The RF spectrum under the condition of $\Delta f_{ceo} = 0$, where $f_{ceo}$ of two combs are both locked at a positive 250Mhz. **b** The RF spectrum in $\Delta f_{ceo} \neq 0$ when one comb's $f_{ceo}$ is locked at a positive 250Mhz and the another at a negative 250Mhz.

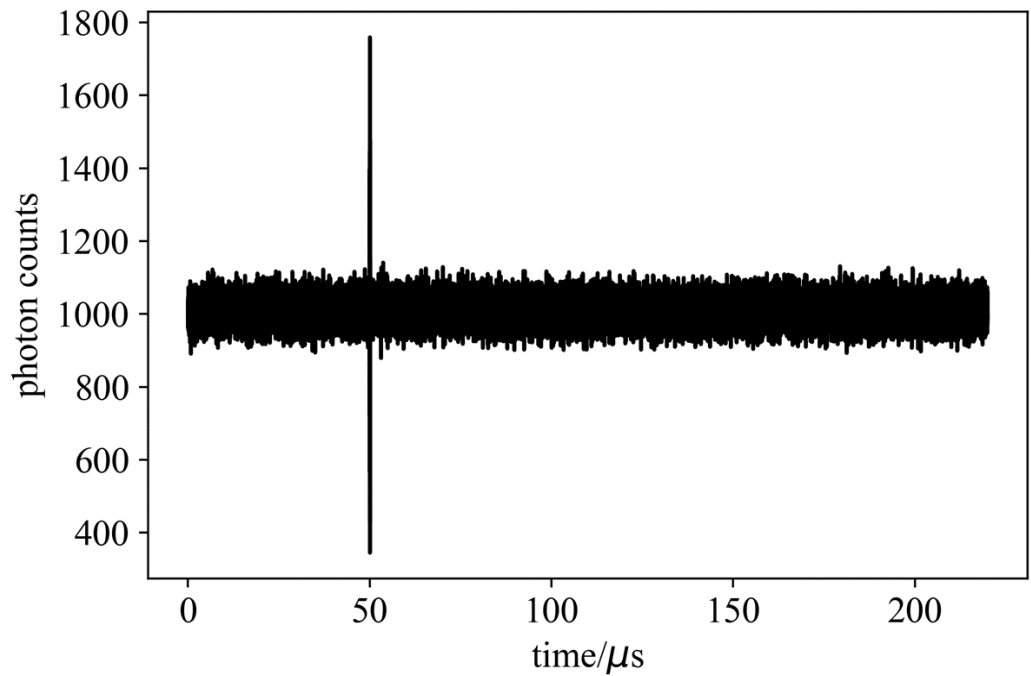

**Fig. 2** The photon-level interferogram after a 300s accumulation, with two combs directly referenced to the Rb atomic clock. The interferogram period is 119.8$\mu s$. Notably, the second central fringe, expected around 170μs in the scan, is absent, indicating a loss of photon counting addition coherence over one interferogram period.





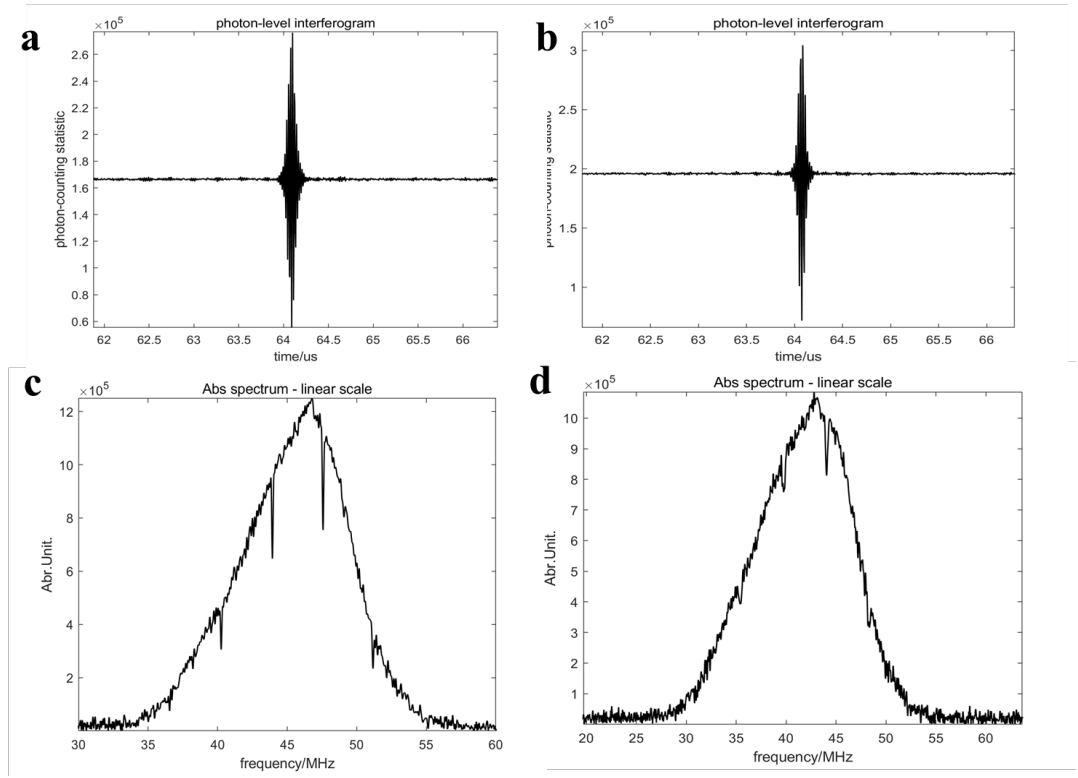

**Fig. 3 a,c** the photon-level interferogram and the RF spectrum with the dual-comb system tightly locked to the optical reference. **b,c** the photon-level interferogram and, and the RF spectrum with the dual-comb system loosely locked to the RF reference (the Rb atomic clock).

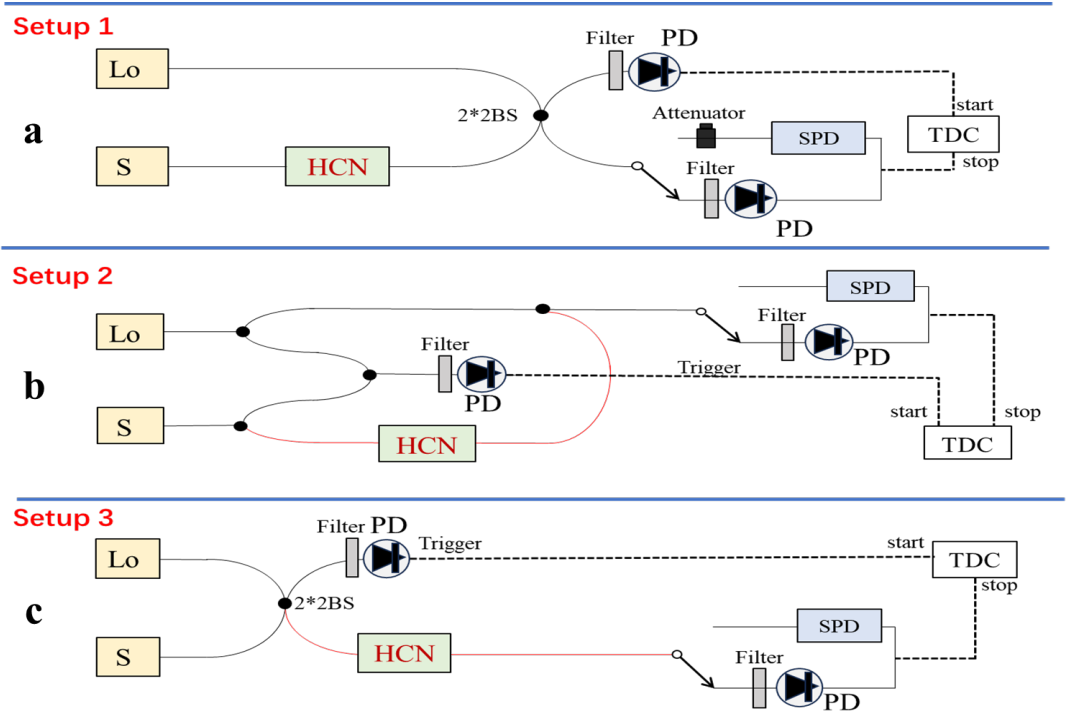

**Fig. 4** Schematic line of 3 types of interferometer setups and the jitter statistic between start signal channel and photon-level DCS sensing channel. The solid line represents optical fibers, and specifically, the red soild lines represents the part that can be used in open-path sensing. The jitter statistic is achieved by substituting the SPD





part with normal PD in traditional DCS detection power and then calculate the time-correlation of the interferograms from the two PDs. The normalized jitter statistic results are shown in Supplementary Fig 5.

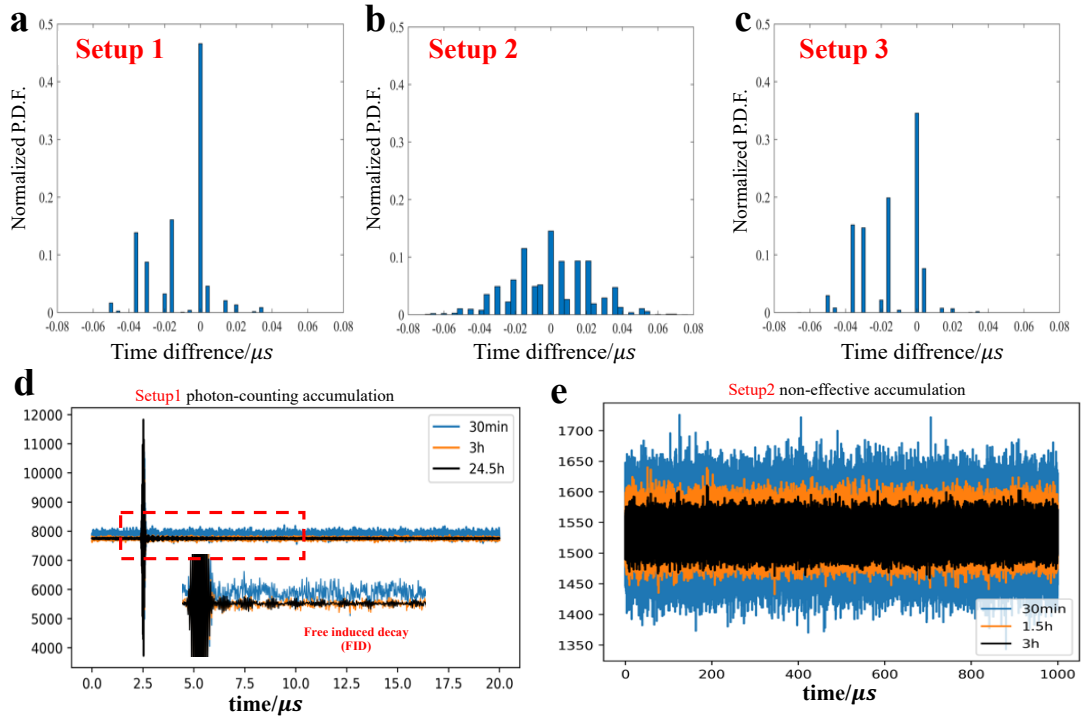

**Fig. 5 a,b,c** Normalized jitter statistic results of 3 types of setups from 1-3 respectively. Details seen in the text. **d** The photon-counting results of setup 1. The interferogram is successfully reconstructed. In this "single probe laser configuration, the desperation spectrum of gas can also be obtained. **e** In setup 2, there is no effective accumulation results due to relative high time jitter between photon clicks and the start signal. Here, the effect of fiber-length wandering is enough to cause the failure of reconstructing the photon-level interferogram.